# Performance metrics in a hybrid MPI-OpenMP based molecular dynamics simulation with short-range interactions

Anirban Pal[1], Abhishek Agarwala[2], Soumyendu Raha[3], Baidurya Bhattacharya[4]*,

[1] Dept of Mechanical Aerospace and Nuclear Engineering, Rensselaer Polytechnic Institute, Troy, NY 12180, USA
2 Archayne Labs, Gurgaon, Haryana 122001, India
[3] Supercomputer Education and Research Centre, Indian Institute of Science, Bangalore 560012, India
[4] Dept of Civil Engineering, Indian Institute of Technology Kharagpur, WB 721302, India
* corresponding author, email: baidurya@civil.iitkgp.ernet.in,
baidurya.bhattacharya@jhu.edu fax: +913222282254

**ABSTRACT**

We discuss the computational bottlenecks in molecular dynamics (MD) and describe the challenges in parallelizing the computation-intensive tasks. We present a hybrid algorithm using MPI (Message Passing Interface) with OpenMP threads for parallelizing a generalized MD computation scheme for systems with short range interatomic interactions. The algorithm is discussed in the context of nano-indentation of Chromium films with carbon indenters using the Embedded Atom Method potential for Cr-Cr interaction and the Morse potential for Cr-C interactions. We study the performance of our algorithm for a range of MPI–thread combinations and find the performance to depend strongly on the computational task and load sharing in the multi-core processor. The algorithm scaled poorly with MPI and our hybrid schemes were observed to outperform the pure message passing scheme, despite utilizing the same number of processors or cores in the cluster. Speed-up achieved by our algorithm compared favourably with that achieved by standard MD packages.

## 1. INTRODUCTION

Multi-core clusters (MCC) have been the standard for computing applications for the last few decades [1, 2]. The multicore processors can be full scale processors, general purpose (GP) and ordinary graphics processing units (GPU) and other processing elements. GPUs [3] often come as assistance to multi-core full scale processors and in more recent times have appeared as the main processing engines which the full scale processors have supplemented with input/output and other functions. An MCC incorporates a two-tier structure: processors across SMP (symmetric multiprocessor) nodes communicate via an external interconnect, and those within the same node are ideally located on the same motherboard and share various levels of cache and memory [2]. Parallel processing on an MCC incorporates a *hybrid* or *mixed-mode* scheme [4, 5] where processors across nodes communicate via MPI, while processors within each node run threads [6], which are essentially instruction sequences within the context of processes. Alternative implementations use threads without MPI on the entire cluster [7, 8]; however this method is not too widely used. Vis-à-vis hybrid schemes, pure MPI implementations on MCCs have proved to be effective [1, 9, 10] as well as ineffective [11-14] and exhibit a cross-over phenomena with data size [4]. Hence, in the present work we explore the hybrid scheme of parallelization over an MCC platform.

There has been considerable work in parallelizing molecular dynamics programs on shared memory space [15-17], as well as distributed memory architectures [18]. NAMD [19], GROMACS [20] and LAMMPS [21] are among the popular molecular dynamics packages which use message passing and threading techniques. Hu et al [22] have proposed efficient MD-EAM algorithms for multi-core platforms. Graphic processing units (GPUs) are being increasingly used to perform MD simulations of larger atomic systems involving millions of atoms and have been known to drastically outperform pure CPU-based implementations [23-26]. They remain a major topic for research.

One way to establish the veracity of molecular simulations is to reproduce well known micro- and macroscopic phenomena. In order to to perform a simulation that can reflect or agree with such phenomena, large regions (up to billions of atoms) may need to be simulated over large time intervals [27]. This imposes a gigantic computational burden. Molecular Dynamics (MD) simulations are highly CPU-intensive and have typically very long running times, which can run into days or weeks. In this context, any significant improvement in the total computational time is desirable. Part of this can be brought about by improving the serial algorithm. For example, the incorporation of "Neighbor Lists" brings down the serial running time from $O(N^2)$ to $O(N)$ for simulations involving only short range interactions [28], (n being the total number of atoms in the system); however such standard techniques are insufficient in bringing down the computation time beyond $O(N \log N)$ when long range (e.g., Coulombic) interactions cannot be ignored and cannot be approximated by short range models (e.g., through Ewald correction). Under such circumstances, where computational efficiency is the topmost priority, large linear gains in computational time can be achieved through parallelization of the existing optimized serial algorithm [29].

The MD algorithm essentially comprises a set of steps that are determined by the integration scheme used to solve the equations of motion of the atoms. If the Verlet integration scheme is used, the entire task could be divided into 3 basic Verlet steps, each step involving the updating of all atomic positions and their derivatives with each time-step

[30]. The parallelization of such an algorithm in a scheme involving both threads and MPI is non-trivial owing to the dissimilar natures of the shared-memory (threads) and distributed memory (MPI) platforms. Threads utilize a common memory space and hence concurrent write operations by parallel threads must be avoided. MPI processes on the other hand have their own private memory locations to write to but must communicate with each other at regular intervals. Unlike processes, threads do not have to communicate with each other. The MD computational task must be first divided among various processes (MPI), and each MPI process then spawns threads to get its assigned work executed in a parallel fashion.

In this paper, we describe an MPI/OpenMP based parallel algorithm in detail for MD of systems with short range interactions and discuss its various implementation and speedup issues; the numerical aspects are demonstrated through a simulation of nano-indentation of a chromium film by a spherical carbon indenter. Chromium and its compounds are used as protective films owing to its corrosion resistance [31] and the dimensions of such films can reach the order of nanometers [32]. The embedded atom method [33] is used to model the interactions among chromium atoms; the interactions between the indenter and the film are modeled by a carbon-chromium Morse-potential [34]. The MD algorithm uses a two-tier Verlet neighbor list [35] to speed up the simulation. The parallelization scheme is implemented on a shared address space computer architecture using multiple threads. Since our usage of threads is limited to within one SMP node only, MPI is used to perform the simulation on a cluster of several SMP nodes in a parallel manner. Thus, MPI is used to communicate between the nodes and threads are used within each nodeto perform the computation. The speed-up and communication overhead obtained by varying the number of SMP nodes and threads used in each system are studied. Speed-up achieved by our algorithm is also compared with that by LAMMPS.

## 2. BACKGROUND

### 2.1 The basic tasks in MD

Any Molecular Dynamics (MD) simulation [30] involves (1) a force computation step based on a relevant potential model, and (2) step(s) where the particle positions and, if necessary, velocities are obtained for the succeeding time-step(s). In this way the atomic system evolves in its phase space from a given set of initial coordinates over a certain period of time, interacting with one or more heat reservoirs if relevant, either toward equilibrium, or away from it if subjected to some external protocol.

We outline the main tasks performed in every time step of the numerical integration scheme in the following, with reference to the nanoindentation process that will form the numerical example later in the paper:

**Task 1**-*Update of neighbor lists*: Based on the assumption that inter-atomic forces (for ground state atoms) in solids become negligible at long ranges [36], atoms are considered to be affected only by its neighboring atoms. In other words, atoms that are far away from each other have a numerically insignificant effect on each other. Thus, when computing the forces on each atom, the effect of far off atoms are ignored and only the neighbors are considered. A set of two lists, an outer dynamic neighbor list [35] and an inner list, which is updated by traversing through the outer list, is used to determine the neighborhood of each atom.

**Task 2**-*Implementation of external protocol*: The hemispherical indenter is moved into the film at a constant velocity, in the form of a displacement boundary condition updated at every time step.

**Task 3**-*Update of atomic positions and velocities*: In this task, the velocity and position of each atom is updated according to the Velocity-Verlet integration scheme [37]. This step uses the position, velocity, and acceleration of each Cr atom at time *t* to compute its position at time *(t + Δt)* and the first corrected velocity at time *(t + 0.5Δt)*.

**Task 4**-*Force & Energy Computation*: The Cr-Cr interaction is modeled using the Embedded Atom Method [38] where the electron density (due to all other neighboring atoms, mentioned in the inner list) at every atomic site is computed first. Then, this electron density is used to compute the force on the corresponding embedded atom. The force between the Cr atom and the indenter atoms is also computed from a C-Cr Morse potential. The total force on (and hence the acceleration of) each Cr atom is thus determined. Phase space functions such as kinetic and potential energies , stresses etc.  can also be determined as part of this task.

**Task 5**-*Update of atomic velocities*: The fifth task of the iteration involves using the updated acceleration of each Cr atom at time *(t + Δt)* evaluated in ***Task 4*** and the velocity at time *(t + 0.5Δt)* evaluated in ***Task 3*** to execute the final step of the Velocity-Verlet integration and compute the final velocity of the Cr atom at time *(t + Δt)*.

All the above tasks must be computed serially one after the other, as is evident by the task dependency graph in the next section. Therefore, there is no scope of parallelization among the various tasks. However, the computations within the various tasks can be parallelized.

**2.2 The parallel computing paradigm & the computer system**

The simulation algorithm incorporates data-parallelism but not process-parallelism as the data produced by one process is consumed by other processes in the succeeding time-steps [29]. This necessitates the sharing of system data among all processes. With data sharing, the problems involved in passing the required data to one process and receiving it from another process are circumvented. These considerations, along with software portability, latency hiding, scheduling, load balancing, and programming ease [39], favor the use of threads over message passing models on the same machine (However, this does not imply that the use of a hybrid scheme which maximizes the usage of threads on the same SMP node will be the best scheme, as is discussed later). Owing to their ease and widespread use, OpenMP threads were selected over other thread paradigms [39]. However, since our usage of threads is limited to individual SMP nodes, MPI must still be used to communicate between the processes running on the different SMP nodes.

The simulations were carried out on a cluster of six Tyrone blade servers, each equipped with two Intel Xeon quad-core processors (thereby 8 cores per blade) with 32KB L1 cache per core, 6MB L2 cache per chip, each chip containing two processors. Communication between the blade servers (SMP nodes) was performed with a 20 Gbps (gigabit per second) *'Infiniband'* switch. This enabled the usage of MPI processes among the SMP nodes or blades and threads among the cores within each blade. The operating system for the cluster was CentOS release 5.2. The code was written in C and was compiled using ICC 11.0 from the Intel OpenMPI 1.6.5 suite with default optimizations (–O2) which provide a good level of optimization, speed and safety.

**2.3 Numerics**

In a distributed memory architecture where MPI is used, numerical accuracy and stability are pressing issues [40]. Owing to finite precision in computer arithmetic, different ordering of computations will lead to slightly different results, the so-called rounding errors. In large scale simulations involving billions of arithmetic operations, numerical consistency is paramount to achieving acceptable results. In an MD simulation incorporating message passing techniques, information such as atomic forces, electron densities, etc. from all the MPI processes need to be added up and communicated back to all the processes. When such summands vary considerably in magnitude (forces at short ranges and short range equivalent of interactions at long ranges), the sum will depend on the order in which the various summands are added [40]. Here we use a self-compensated summation method (SCS) [39] in order to minimize round-off errors and achieve consistency and accuracy.

**2.4 The Inter-atomic potentials**

The single most essential feature which determines the quality of the simulation is the model used to describe the inter-atomic forces. The atomistic simulation of the indentation process requires suitable potentials to model the chromium atoms (film) and carbon atoms (indenter). The metallic bonding in Chromium requires the use of the Embedded-Atom-Method (EAM) which incorporates an approximation to the many-atom interactions neglected by pair-potential models [41]. This method developed by Daw and Baskes [33] has been improved and modified by Yifang et al [42], to fit the negative Cauchy pressure of Chromium and has been adopted here. The indenter is hemispherical in shape and is composed of fixed carbon atoms in order to simulate the hardness of diamond. The carbon-chromium (C-Cr) interactions have been modeled with Morse pair potential taken from [34]. Details of the potentials have been given in the Appendix. For the sake of portability and transparency but at a cost of increased computational time, we have not tabulated the interatomic potentials and forces, but have used the analytical forms unlike commercial/ shareware MD packages.

## 2.5 System initialization and loading

We looked at various sizes of the BCC Chromium film: ranging from 10x10x10 unit cells (2331 atoms) to 40x40x40 unit cells (132921 atoms). Figure 1 shows a typical film: 19 unit cells each in the x and y directions and 14 unit cells in the z direction, i.e., 61x61x45 Å$^3$ consisting of 11054 atoms. The indenter is a moving rigid hemispherical displacement boundary composed of 394 C atoms. Free boundary conditions were applied to all the vertical faces and the bottom-most layer of unit cells was kept fixed. The initial temperature was 10K and the velocities of each atom were sampled randomly from the corresponding Gaussian distribution. No thermostat was used and the indenter was kept rigid to simulate the hardness of diamond. The time-step was 1 femto-second. After the system was initialized, the velocities were corrected to make sure that the centre of mass of the whole system remained stationary.

It is important to note that setting boundary conditions is an essential feature of MD simulations and the results obtained may heavily depend on it. On a given surface, a set of traction-free boundary conditions would imply the plane stress condition while a set of fixed displacement boundary conditions would imply the plane strain condition. If a dynamic process such as indentation is being studied, boundary conditions such as "slab" periodic boundary conditions or fixed boundary conditions would limit the vibration modes (frequencies) attainable by the system [43] and could compromise the integrity of the observed phenomena. In such cases, one should ideally resort to multi-scale methods [44] in order to apply suitable boundary conditions via continuum models such as finite elements.
For performance studies and comparison with LAMMPS, the simulation was run for 300 steps without any indenter motion. For simulating the complete nanoindentation process, the Chromium atoms were initially allowed to relax for 10,000 timesteps, following which the indenter was moved in with a speed of 0.0001 Å/fs (=10 m/s) up to a depth of 3.08 Å and then withdrawn at the same speed. The load on the indenter was measured by summing up the forces on the indenter in the direction of indentation ($z$ direction), and the displacement of the lower-most atom of the indenter from the initial free surface was taken as the depth of indentation.

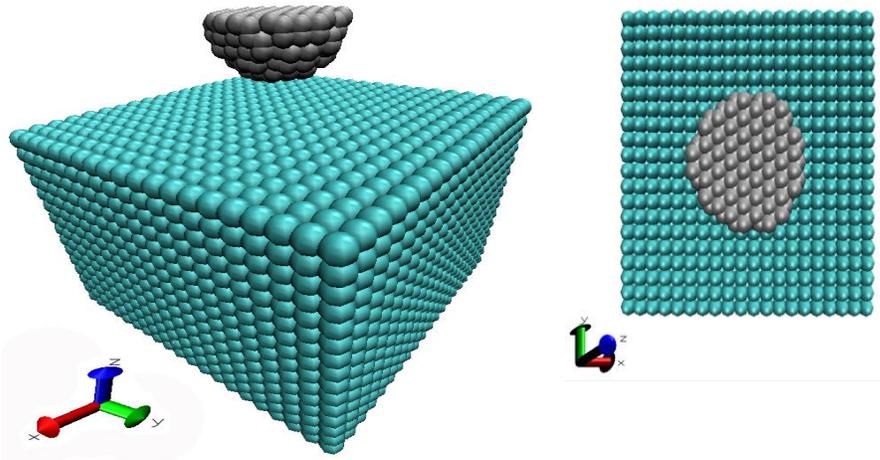

**Figure 1: (left) Isometric and (right) top views of the indentation setup**

### 2.6 Two step neighbor lists

To reduce the computational time, a set of two neighbor lists was used for every Chromium atom. In solids, the atoms vibrate about their mean positions and hence the larger list (outer list) needs to be updated only rarely compared to the smaller (inner) list. The outer list for each atom stored a larger neighborhood of atoms and was dynamically updated, i.e., the outer list was updated whenever any atom got displaced by more than half the skin thickness. The inner list stored all the neighboring atoms of every atom and all force computations were done with respect to this list. This inner list was updated using atoms from the outer list (**Figure 2**). Alternatively, one may use a cell-linked list or a Verlet cell-linked list in which the whole domain is divided into cubical cells, where the neighborhood of an atom is determined by the cells adjoining its host cell. The reader is referred to [45] for a discussion on this topic.

The "outer list" was created for each atom corresponding to a radial distance of $1.7r^*$ where $r^* = 8.262$ Å is the cut-off distance taken to be thrice the equilibrium inter-atomic Cr-Cr distance (2.754 Å [46]). The skin thickness governing the updating frequency of the outer list was taken to be $0.7r^*$. The inner neighbor list was created for each atom from its outer list with a distance of $r^*$. The forces/accelerations on all the atoms were calculated considering the atoms only in the inner list. The inner list was updated every time-step. Since in an update operation each atom in the list is visited exactly once during which the neighborlist, a set of constant cardinality independent of the number of atoms is traversed, the computational complexity of the procedure is linear in time.

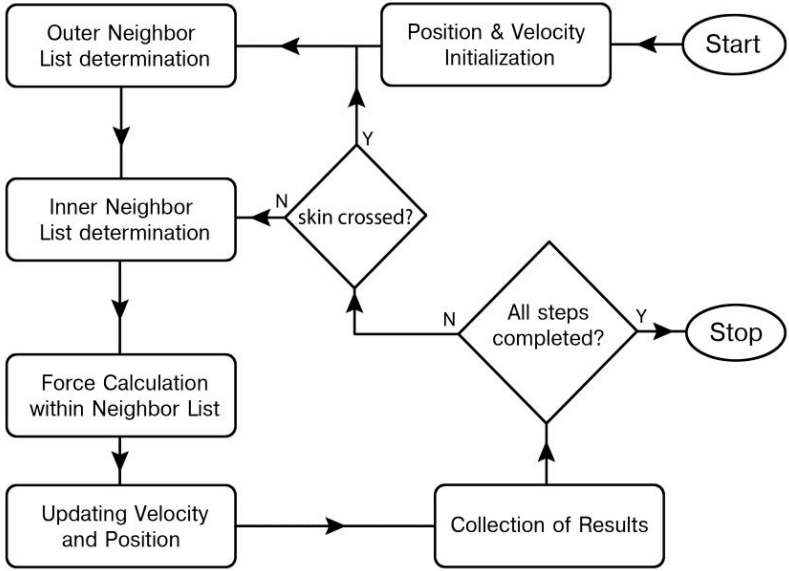 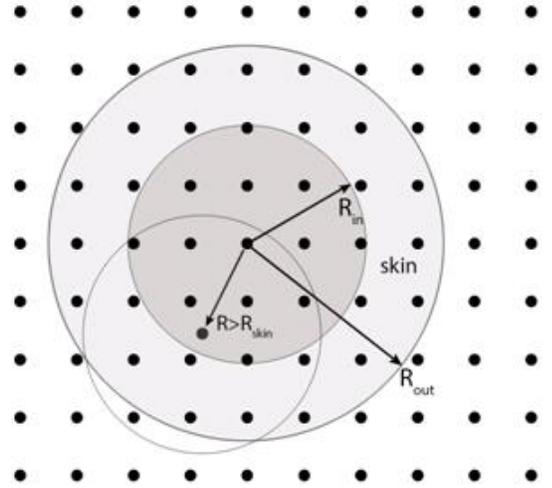

**Figure 2: Left: A flowchart depicting the various steps of the serial algorithm. Right: the two step neighbor list: when the central atom moves by a distance larger than the skin distance, the outer list needs to be updated**

## 3. PARALLELIZATION

### 3.1 What to parallelize?

In any MD simulation, the force and energy computation steps as well as the neighbor list creation steps are the most CPU-intensive operations [47]. The embedded atom method poses the additional task of computing the electron densities at all atomic sites. The entire task of calculating the forces on all the atoms is divided equally among all the processors, i.e. if the total number of available processors is P and the number of atoms N, then each processor is assigned the task of computing on N/P atoms. This accomplishes a linear reduction in the running time of the algorithm, which is ideally proportional to the number of processors used. However, the actual gain achieved is sub-linear, due to overheads involved in the parallelization routine and the non-parallelizable part of the algorithm.

The electron density $\rho_i^t$ at each atomic site $i$ is the sum of pair-wise electron density functions $f(r_{ij}^t)$ of all its neighbors $j$. The force on each atom $F_i^t$ is then a weighted sum of the function derivatives ($F'$) of these electron densities and the two-body forces $V'$ [48], where primes (′) denote differentiation, $q_i^t$ refers to the position vector of atom $i$ for timestep $t$ and $r_{ij}^t$ is the distance between atoms $i$ and $j$.

$$F_i^t = -\sum_j \left[\left(F'(\rho_i^t) + F'(\rho_j^t)\right) f'(r_{ij}^t) + V'(r_{ij}^t)\right] \frac{q_i^t - q_j^t}{|r_{ij}^t|} \tag{1}$$

The details of the potential are given in the Appendix.

The total computational time of the MD simulation is determined almost entirely by the iterative loop running thousands of times that performs the neighbor list computation along with the electron density and force computations. Two 3xN arrays store the velocities and accelerations of all the N atoms. A 5xN array stores the positions as well as values of two electron density functional derivatives required for force computations. Every atom of the film is also associated with a data structure containing the two neighbor lists (an outer list and an inner list). Such structuring makes the inter-process communication transparent and easy to implement with the existing MPI collective-communication functions.

A relatively small amount of resources are spent on (i) the initialization stage where the velocities and the positions of the atoms of the system are initialized, (ii) computation of time histories of derived quantities such as stresses and temperature if necessary, and (iii) the final stage where some equilibrium properties may be calculated. Hence, one needs to look at the task dependencies among the various stages of computations taking place inside the iterative loop in order to work out a successful parallelization scheme.

**Table 1: Task list during MD iteration**

| Task   | Description                                    |
|--------|------------------------------------------------|
| Task 1 | Update of neighbor lists                       |
| Task 2 | Indenter motion                                |
| Task 3 | Atomic positions and velocities updated        |
| Task 4 | Electron density, force and energy computation |
| Task 5 | Atomic velocities updated                      |

The task dependency graph (Figure 3**Error! Reference source not found.**) illustrates the interrelationships between the five tasks (listed in Table 1) into which the iterative loop has been divided. As evident, other than tasks 1 and 2, no two tasks can be executed in parallel due to the nature of their interdependencies. In other words, tasks 1 and 2 must be completed before initialization of task 3 with tasks 4 and 5 following sequentially after task 3 is completed. However, each task can be parallelized within itself: the computational load associated with each task can be divided among the various processors by a suitable decomposition technique (Figure 4). Although the use of force decomposition and spatial decomposition methods is prevalent [18], we employ the atom decomposition technique forits simplicity. The force decomposition method computes and divides a force-matrix block-wise among the processors and offers better scaling ($O(N/\sqrt{P})$) over atom-decomposition ($O(N)$). Spatial decomposition offers even better scaling ($O(N/P)$) if the geometry can be optimally divided into processors, but is much harder to implement [29].

However, we do not use force and spatial decomposition because of the significant overhead incurred in the present application to keep tab of atoms leaving and entering the domain due to the flow like behavior of the material in the immediate neighborhood of the indenter. Also, the algorithm incorporates "all to all" communication and will not benefit significantly from a better decomposition strategy.

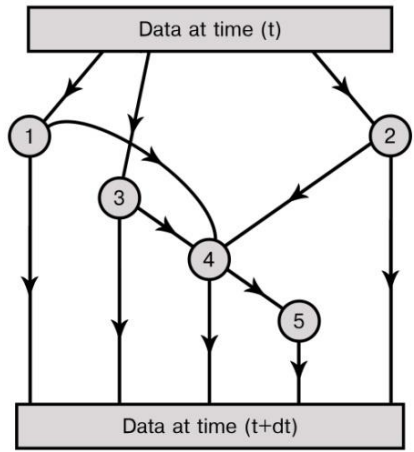 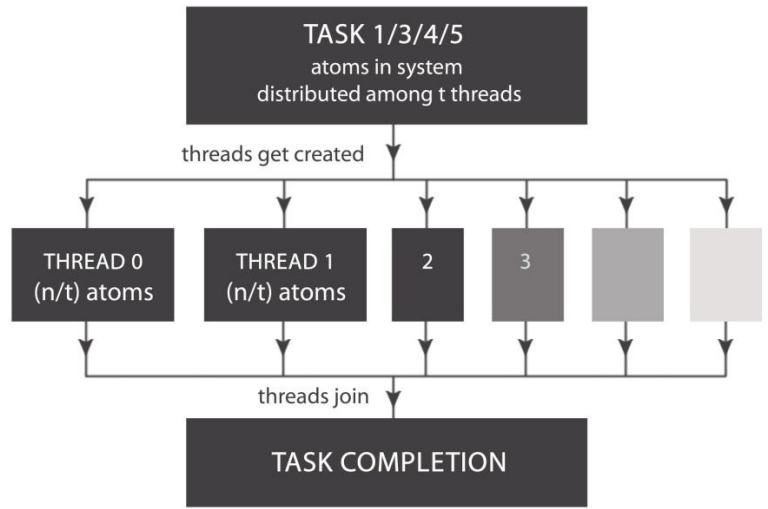

**Figure 3:: The task dependency graph of the various tasks of the algorithm.**

**Figure 4: Tasks 1,3, 4 and 5 can be parallelized among themselves**

### 3.2 Parallelization Issues

Within each MPI process, parallelization is implemented using threads on a "shared address space" platform consisting of eight processors. The "shared-address-space" parallel platforms have a common data space that is accessible to all the processors. This enables simultaneous read operations to be carried out on such a platform [39]. But concurrent write operations must be properly scheduled by the programmer such that no two threads try to write to the same memory location. Critical sections can be used but being highly expensive they are discouraged [49].

The tasks that require threading are the inner list update, the electron-density computation, and the force computation. Each of the above steps comprises an outer-loop which runs through all the atoms, and an inner loop which runs through the appropriate neighbors. Only the film-indenter force computation step involves simultaneous write operations to both the inner loop and the outer loop indices, and hence must be parallelized carefully. The remaining steps involve write operations only to the outer loop index and hence can be parallelized by distributing the outer loop among various threads, as explained in section 3.3.1.3.

The division of the computational load among various MPI processes requires the appropriate inter-process communication at necessary points in the program. In order to obtain maximum efficiency, the communication must be minimized. This can be achieved by efficient partitioning of the atomic system.

### 3.3 Programming Technique

Computer hardware architecture has a crucial role to play in the performance aspects of the parallelization scheme. The size and manner in which caches are shared among cores [52, 53], and interconnect bandwidth [50, 51] and latencies [52,54,55] determine speed-up and scalability. However, these issues are beyond the scope of this study.

```
Initialize MPI WORLD
Initialize variables and allocate memories
Divide system for MPI
Initialize the film and indenter and assign velocities as per temperature
MPI BCAST all relevant variables ( position, velocity & acceleration)
###Start LOOP iterating over time steps
  update outer lists if required (MPI_reduce distance traveled flag)
  Make inner list
  Move indenter
  ##Compute forces (verlet steps)
    Verlet Step 1
    Calculate electron densities at atom site
    MPI allgather of electron densities
    compute force on atom
    MPI allreduce (3PotEs and KinE)
    Verlet step 3
  ##END Compute forces
  MPI Barrier
  MPI gatherv to gather all data into root for processing physical
properties
  Output result files at regular intervals
###End LOOP
Finalize
```

**Figure 5: Pseudo-code describing the MPI/OpenMP parallelization of the MD simulation of nano-indentation**

A pseudo-code describing our algorithm is presented in Figure 5. All loops that step through individual atoms and are data parallel (eg. verlet step 1) are encapsulated in an "openmp parallel for" statement to achieve thread level parallelization. The advantage of using multiple threads on a single computer system (SMP node) is well known and this has been coupled with the idea of using several nodes in a parallel manner. The Message Passing Interface (MPI) has been used in conjunction with OpenMP [17] to perform the simulation. The implementation in this paper is fundamentally similar to the 'Hybrid masteronly' model [10], which uses one MPI process per node and OpenMP on the cores of the node, with no MPI calls inside the OpenMP parallel regions. POSIX threads or pthreads is a low-level API, and offers more flexibility but less workability than OpenMP which is simple to implement. One would use pthreads specifically in applications which require fine control over thread management. However, since the current algorithm was trivially parallelizable, the benefits of using OpenMP were more pronounced.

Performance analysis tools such as EXPERT [56], Vampir [57] could be used to monitor MPI as well as OpenMP performance metrics, and subsequently optimize the application. The presence of inactive threads and under-saturation of inter-connect bandwidth can reduce performance of hybrid schemes [9]. Additionally, the level of threading support [10], i.e., single, funneled, serialized or multiple can allow the programmer to experiment with alternative algorithms.

*3.3.1 Implementation: Threads*

As stated previously, threads are used within each MPI process. Data-parallelism is employed to divide the process task among the threads. The difference between partitioning among processes and among threads is that each MPI process has a private memory space whereas all threads of a particular process share the same memory space. As a result, threads do not have to communicate with each other as processes do. Since they share the memory space, mutex locks need to be used to allow threads to write to the same memory location. However, the usage of mutex locks has been avoided by circumventing synchronous write operations on the same memory location, as explained in section 3.3.1.3.

3.3.1.1 Creating neighbor lists

The inner neighbor list computation (done every step) involves the use of a nested loop where the outer loop runs over all the atoms with an inner loop running over the indices of each atom's outer list of neighbors. This procedure involves writing of data corresponding to the atom having the outer loop index and hence the outer loop can be parallelized. Since the outer neighbor list is updated dynamically, it happens infrequently and need not be parallelized.

One may use a dynamic linked list (DLL) or a static array (SA) of a suitably predefined size to store the outer and inner neighbor lists. The functions associated with the creation of neighbor lists using DLL involve memory allocation operations and hence cannot be threaded with conventional memory allocation functions (calloc(), malloc()) which are non-reentrant. Non-reentrant functions are those functions which when called simultaneously from different processes or threads will not return different values. In other words, they cannot be called from more than one thread at a time. Threading can thus lead to memory management errors and/or yield flawed results in this case. The absence of reliable thread-safe memory allocators imposes restrictions on parallelizing this step. This makes the parallelization of a DLL based implementation difficult. An SA based implementation is thus preferred here owing to its simplicity in spite of its larger memory costs. Two large arrays of sizes defined by the maximum possible number of neighbors are initially created and are used to store the inner and outer neighbor lists correspondingly.

3.3.1.2 Computing functional derivatives at atomic sites

As can be seen from Eq.(1), net force depends on the density terms only via their embedding function derivatives. Therefore, in the nested loop, the net electron density at an atomic site is computed by summing the pair-wise contributions from all its neighbors and simultaneously the functional derivatives are calculated and stored. Since the loop involves computations and write operations for one particular 'outer-loop' atom, the outer loop can be parallelized as there are no simultaneous write operations. The only issue that needs to be considered while parallelizing using OpenMP is the declaration of private variables.

3.3.1.3 Computing forces on each atom

This step involves the calculation of the net force on each atom as per Eq.(1). Again, only the atom corresponding to the outer-loop experiences write-operations, even though read-operations are simultaneous. This is also parallelized in a similar fashion as in computing functional derivatives.

The forces between the film atoms and all the indenter atoms must also be computed. However, each computation is associated with write operations to both the outer loop (film atoms) and inner loop (indenter atoms) indices. If the write operations are not properly scheduled, race conditions will occur. To avoid that we use a cyclic pairing scheme so that the inner loop can be threaded without the need for locks as is explained in Figure 6. Pairs are considered between the outer ring of $N$ film atoms and $iN$ indenter atoms. First $iN$ pairs are formed between the indenter atoms and film atoms from 1 to $iN$. Next, another $iN$ pairs are formed with the film atoms from 2 to ($iN$+1). In this way, over all the selections, all possible pairs are considered and in each selection, independent pairs are selected. The pairs belonging to each selection can be distributed among threads, thereby avoiding race conditions.

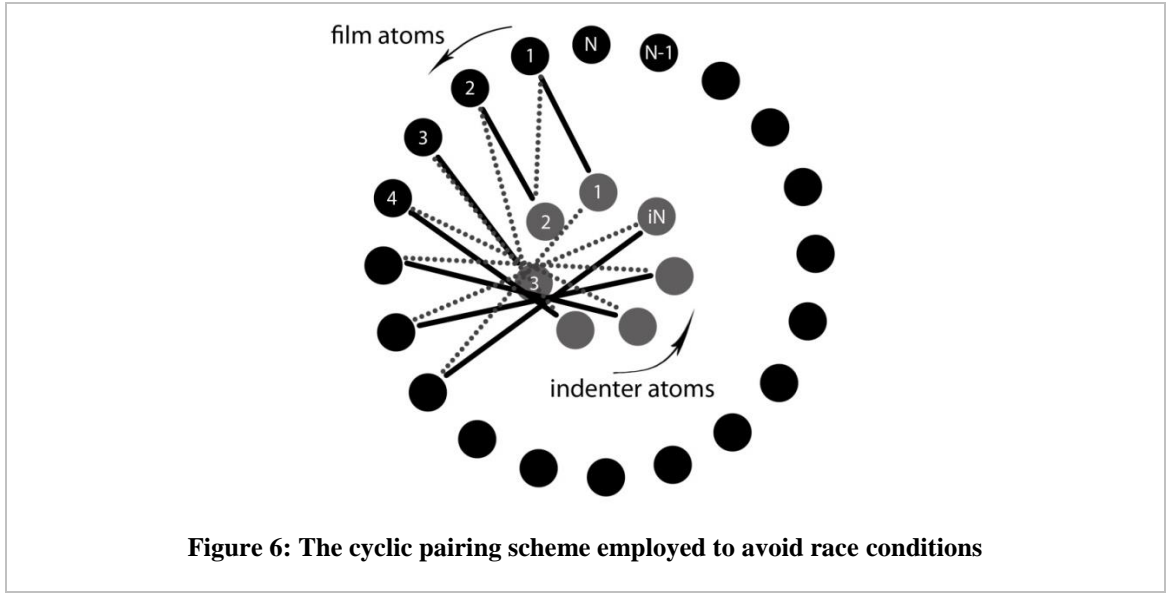

**Figure 6: The cyclic pairing scheme employed to avoid race conditions**

*3.3.2 Implementation: MPI*

The entire assembly of atoms is first partitioned and each partition is assigned to a process as shown in Figure 7. Thus, each process 'knows' the atoms that it must work on. Once the positions and its derivatives are initialized or read from a file in the root process, they are broadcast *(MPI_Bcast)* to the various processes.

Next, each time step of the computation is associated with a set of necessary inter-process communication directives:

    Communication (MPI_Allgatherv) of the updated atomic positions from each process to all other processes after Task 3 (Figure 8).

    Communication (MPI_Allgatherv) of the functional derivatives of the embedding function and modified terms associated with each atomic site from each process to all other processes before the force computation step in Task 4.

    Reduction or sum (MPI_Allreduce) of energies (potential and kinetic) computed by each process and the sum communicated to all processes after Task 4.

    Reduction (MPI_Reduce) of the forces on the indenter atoms computed by all the processes after Task 4

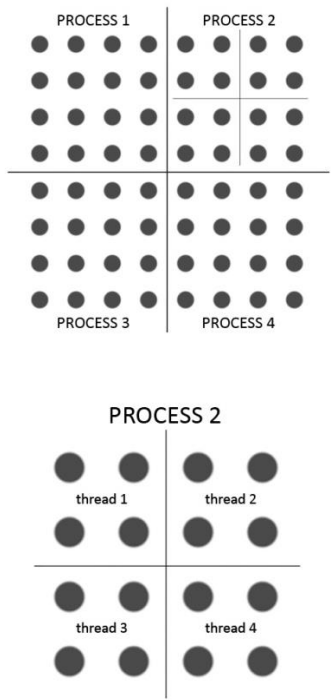 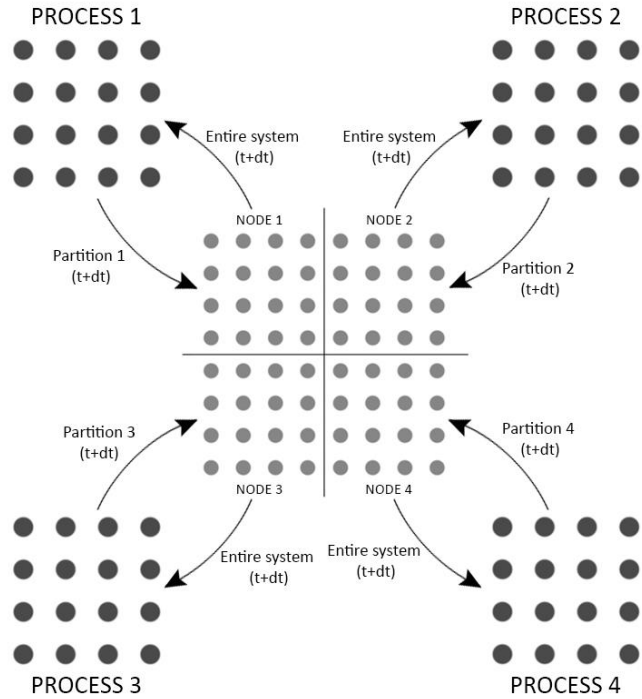

**Figure 7**: Partitioning scheme     **Figure 8**: Gathering updated positions after Task 3

Since each process computes the total force (and the accelerations) on the atoms under its purview, it does not need to gather additional information on the forces on its atoms from the other processes, as the other processes are concerned with an entirely different set of atoms. Hence the communication of the forces on each atom is replaced by the communication of the embedding function derivatives at each atomic site.

## 4. NUMERICAL RESULTS AND OBSERVATIONS

Our focus in this section is to study the performance of our hybrid MPI/OpenMP parallelization scheme, rather than the actual indentation process itself. An excellent review of simulation of nanoindentation for the interested reader may be found in [58]. We looked at various sizes of the BCC Chromium film: ranging from 10x10x10 unit cells (2321 atoms) to 40x40x40 unit cells (132921 atoms). The indenter was a rigid hemispherical displacement boundary of C atoms in each case. Figure 9- Figure 13 depict the performance of various parallelization schemes (including LAMMPS) in which one or both of the following systems were studied for 300 time steps without indenter motion: (i) The **11k system**: a Cr film of 19x19x14 unit cells (11054 Cr atoms and 394 C atoms) and (ii) The **100k system**: a Cr film of 40x40x30 unit cells (100111 Cr atoms and 3173 C atoms). Figure 14 describes the complete nano-indentation of the 19x19x14 unit cell system using our hybrid algorithm performed in 300000 time steps. The performance metric of the various parallelization schemes are discussed next.

**4.1 Unthreaded MPI vs. hybrid MP/OpenMP schemes**

The MPI process distribution among the 48 different cores in the 6 SMP nodes in our computer cluster was done automatically by the Message Passing Daemon (MPD), such that each node runs a nearly equal number of processes. Figure 9 depicts the speed-up (defined as the ratio of time taken with 1 unthreaded process to time taken with $n$ unthreaded processes) obtained by using an increasing number of MPI processes. For the 11k system, the speed-up flattens after only 18 processes have been put to use: from this point on, adding more processes up to the maximum of 48 in the cluster does not yield much greater speed-up (poor scalability[12]). None of these processes employed any threads in their execution. Such a limit is reached owing to the fact that the effectiveness of using more MPI processes is eventually offset by the inter-process communication overheads. For the larger system however, scaling is linear up to 24 MPI processes and flattens beyond that. This can be explained by the relative size of the communication overhead versus computational task per processor. The communication load increases with increasing number of processors, while the computational task (per processor) decreases for the same system size. Hence, better scaling is achieved if the computational task is a larger fraction of the overall computational burden.

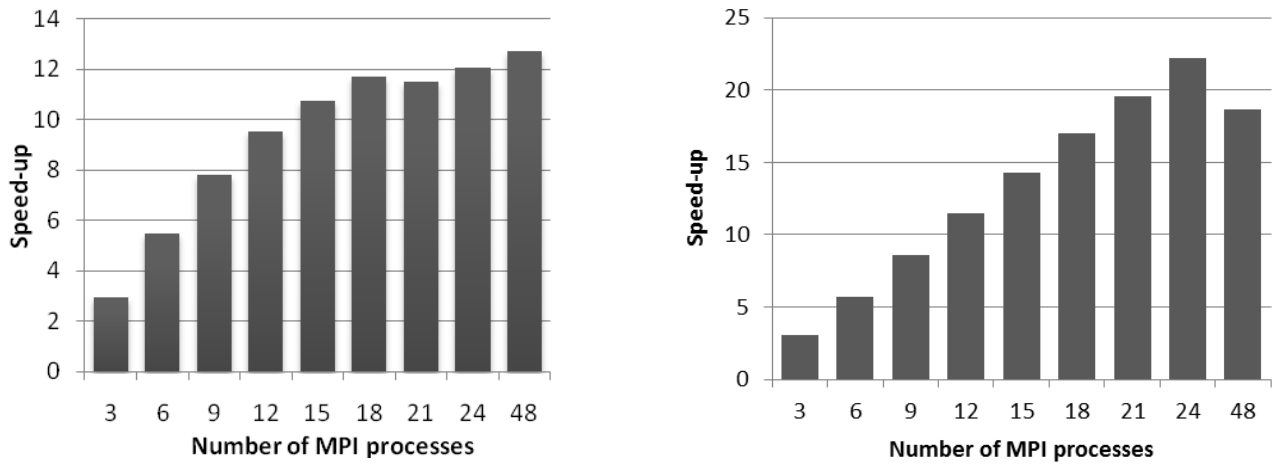

**Figure 9: Speed-up obtained with increasing number of MPI processes (unthreaded). Left: 11k system. Right: 100k system.**

We now introduce threads in our computations. Although there is no limit to how many threads a process can spawn, we looked at up to 8 threads per MPI process as there were 8 cores per SMP node. Figure 10 shows the speed-up for various parallelization schemes starting from the simplest serial execution (1 MPI Process × 1thread/process) up to using the full capacity of the computing cluster (6 MPI Processes × 8threads/process) with each SMP node running a maximum of 1 process, as determined by the MPD. For the 11k system, in the 6x8 scheme, the speed-up obtained is 22.4, which is significantly larger than the speed-up of 12.7 obtained by using 48 unthreaded MPI processes (Figure 9). For the 100k system, the speed-up achieved is even higher: about 41 for the 6x8 combination. In addition, the scaling for the 11k system was sublinear beyond 4 threads; for the 100k system linear scaling up to 8 thread was observed, suggesting good scalability for the algorithm. This achievement is an outcome of the shared memory programming platform on which threads run, where inter-thread communication is avoided. Hence the communication overheads are due to the 6 MPI processes, but the productivity is that of all the 48 processors in the cluster.

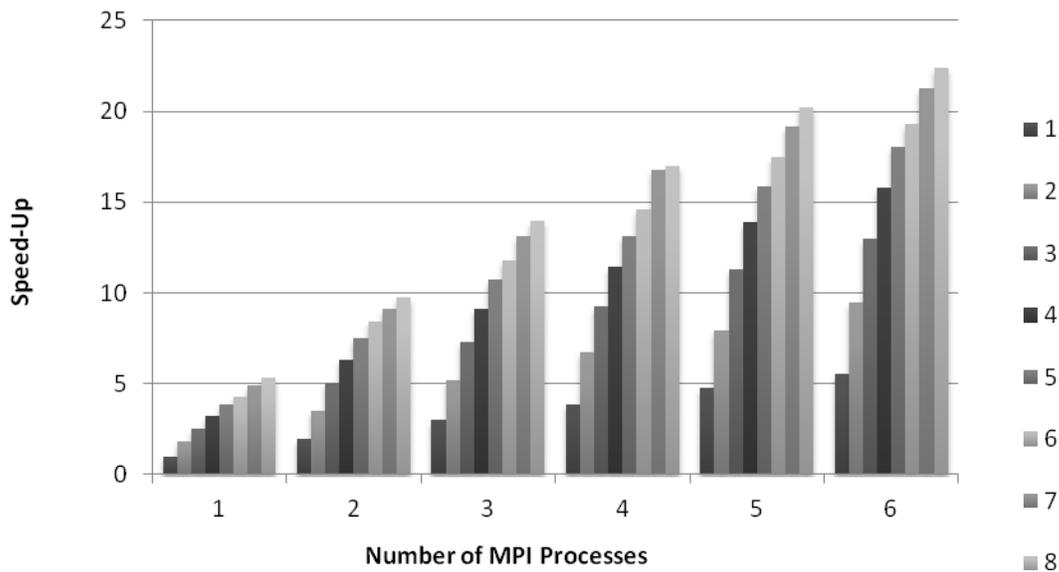

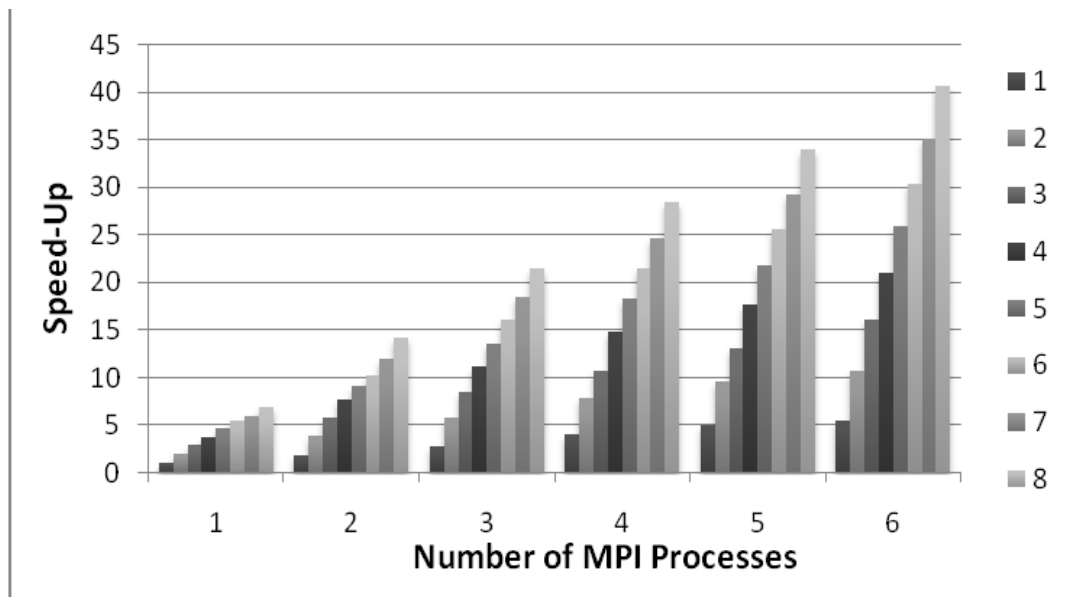

**Figure 10: Speed-up using different numbers of threads (shades of grey) and different number of processes (top) 11 k system (bottom) 100 k system**

Figure 11 is an illustration of the time taken for inter-process communication (IPC) in various process-thread configurations. The IPC time was determined by the total time taken by the MPI directives of Section 3.3.2 to execute. It can be seen that for a fixed number of MPI processes, the overhead does not vary significantly with the number of threads spawned by it and ideally it should remain constant. The overhead however increases with the number of MPI processes used, which also forms the central reason for the speed-up limit as was observed in Figure 9.

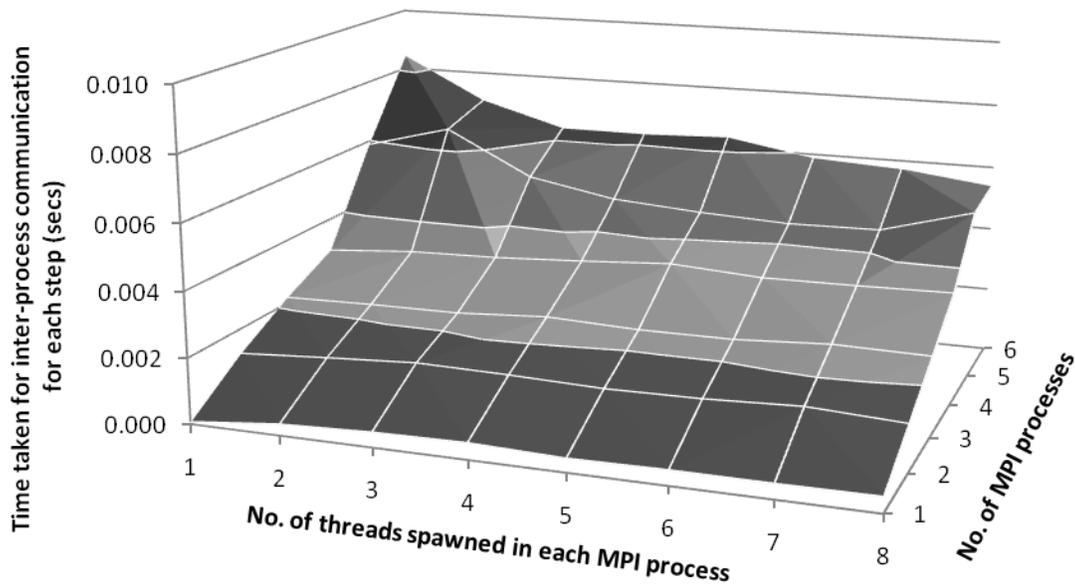

**Figure 11: Inter-process communication overhead in various parallelization schemes (11k system)**

A comparison of various parallelization schemes would be useful here. It was seen that the speed-up limit is not yet reached with 12 unthreaded MPI processes (Figure 9). Hence one could use 12 MPI processes and spawn 4 threads in each process to perform the computation (in order to utilize all the processors in the cluster). A comparison of such competing schemes is depicted in Figure 12 **Error! Reference source not found.**. The 6x8 and the 12x4 schemes work comparatively well, with a 6x8 scheme beating the 12x4 scheme by a small margin. The slowdown in the 24x2 scheme is possibly correlated to the fact that unthreaded MPI scalability gets saturated at 18 processes (Figure 9). The unthreaded 48 process scheme takes the longest time owing to poor scalability as discussed earlier.

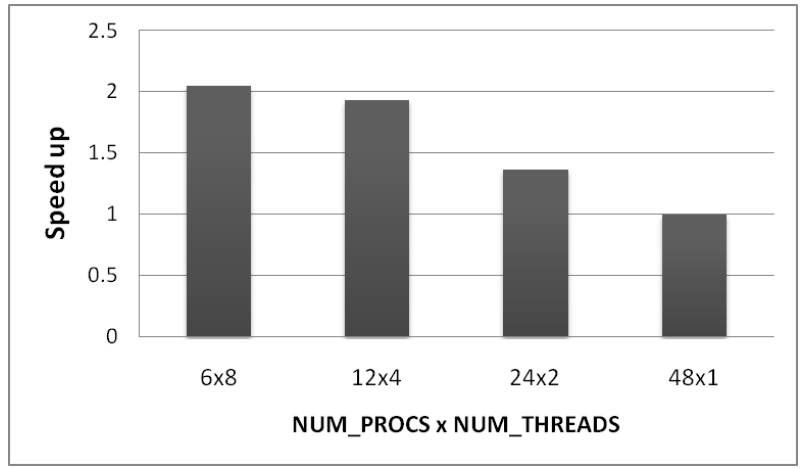

**Figure 12: Time taken in various parallelization schemes (11k system)**

The scalability of the various competitive parallelization schemes, i.e. 6x8, 12x4, and 24x2 with increasing system size is described in Table 2. The system size corresponds to the film size and the indenter interactions were disabled as they contribute only a linear term (~N) to the time taken. The 6x8 configuration was fastest for small system sizes (small no. of atoms), but for larger systems the 12x4 configuration gave markedly superior performance. This could be attributed to the memory architecture (cache sharing) of the dual quad-core processor and the Interconnect latencies. In our system, as mentioned earlier, the 2 cores in each chip shared the same 6 MB L2 cache. Thus, the performance of the Inter-connect vis-à-vis intra-node communication could favor a particular hybrid scheme for a particular computational load.

**Table 2 Scalability for various parallelization schemes (average time/step in seconds)**

|  | System size >> (number of unit cells in x,y,z directions) | | |
|---|---|---|---|
| **MPI x thread** | **10x10x10** | **20x20x20** | **40x40x40** |
| 6x8 | 0.0077 | 0.0575 | 0.4523 |
| 12x4 | 0.0094 | 0.0567 | 0.4110 |
| 24x2 | 0.0166 | 0.0749 | 0.4686 |

## 4.2 Comparison with LAMMPS

We now study the speed-up achieved by the general purpose MD package LAMMPS[21] when simulating the same system under identical conditions. We adopt the 100k system with no thermostatting, timestep of 1 fs, and study its evolution for 300 timesteps as above. LAMMPS runs in a significantly shorter time than our code because the interatomic forces are read and interpolated from a table in LAMMPS whereas in our code, forces are computed analytically in every time step as given in the Appendix and thus take the largest share of time when executing our code. Nevertheless, as stated earlier, we chose analytical computation of forces to ensure transparency and portability. Also, the decomposition method used by LAMMPS (spatial decomposition) offers much better performance for large systems.

The speed-up achieved by LAMMPS when using up to 48 MPI processes (i.e., using full resources of the cluster) is between 8 and 12 and is rather noisy. In comparison, our best speed-up for the 100k system is about 41 which is larger than that from LAMMPS.

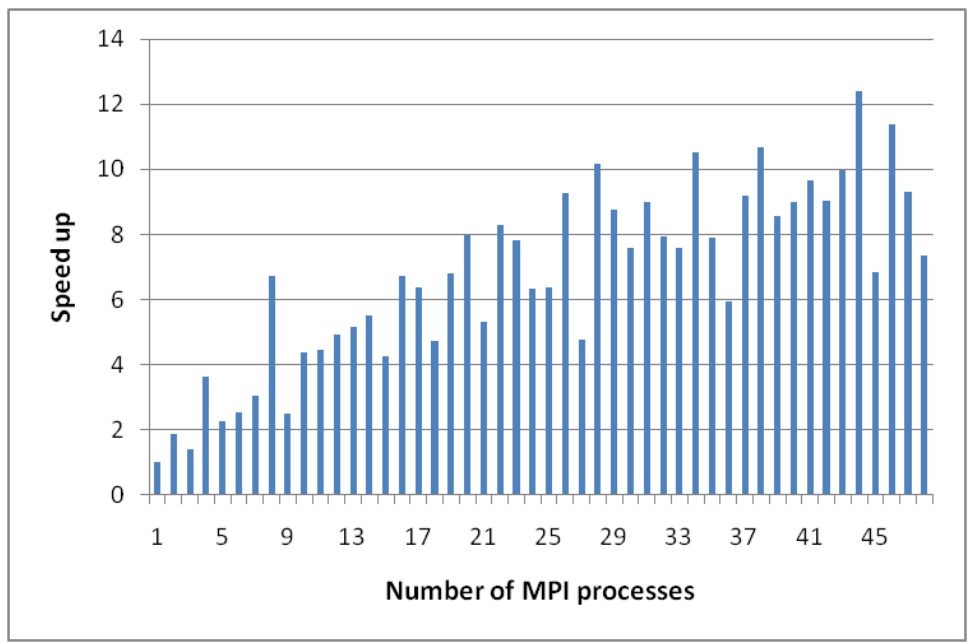

**Figure 13: LAMMPS performance for 100k system**

### 4.3 Simulation of complete nano-indentation

Finally, we look at the load-displacement curve from the nano-indentation simulation of the chromium film (Figure 14). The film was initially allowed to relax for 10,000 time-steps and then the indenter was moved 0.0001 angstroms into the film with every time-step (equivalent speed of $10^{-4}$ Å/fs =10 m/s). The entire simulation was run for 300,000 steps during which the indenter was moved in by 3.08 Å and then taken out of the film at the same speed. The optimal 6x8 process-thread scheme was used and the entire simulation took 19,600 CPU seconds (real-time) to complete.

Although we are using an atomistic approach, we adopt Hertzian contact mechanics to compute elastic properties of the Chromium film. We take the input parameters for the calculations as follows: indenter radius $R = 9.78$ Å, depth of indentation $d = 3.08$ Å, maximum load $P = 27.4$ eV/Å, initial slope of unloading curve (Figure 14) $dP/dh = 23.12$ eV/Å$^2$. Using the Oliver & Pharr method described in [59],we get the composite modulus of the film-indenter system (E*) as 265.7 GPa. The modulus of chromium then comes out to be 321 GPa which is close to that of sputtered chromium films (285 GPa [60]). The contact radius is computed to be 7.1 Å, which gives the hardness as 27.5 GPa which is comparable to 21.61 GPa [61] observed for 400 nm thick chromium coatings.

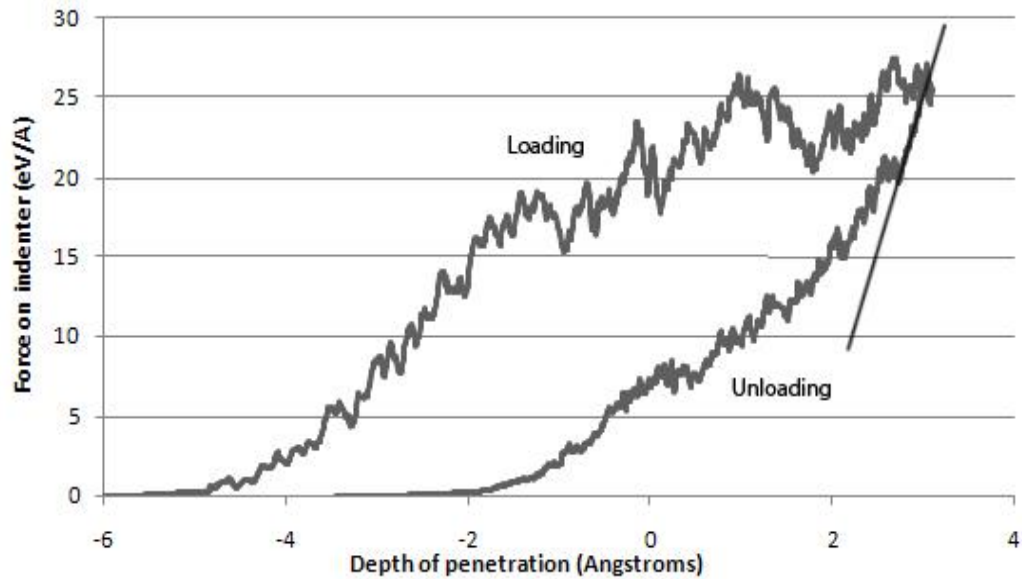

**Figure 14: Load Displacement curve for spherical indenter of the 11k system with speed of 0.0001 A/fs (10m/s)**

## 7. CONCLUSIONS

We have developed a computational scheme for MD simulations that exploits thread-parallelism as well as message passing techniques and implemented it on a cluster of 6 dual-quad-core blade servers (SMP nodes), connected using Infiniband. The challenges and issues of such schemes were discussed in detail. We have shown that such a coupled scheme can work nearly twice as fast as a pure message-passing based implementation for certain system sizes, owing to the additional overheads in the latter being circumvented by the former scheme. For larger systems, however, with increasing work load per SMP node, the performance differential may become negligible. When using unthreaded MPI processes with this algorithm, the speed-up obtained saturated quickly on the cluster, well before the total number of available cores were utilized.

A set of hybrid schemes were compared and were found to be competitive. On larger clusters, there might be several such schemes and their performance will depend heavily on the processor (core) architecture. Certain code-optimizations and computational loads may favor one particular scheme over the other and hence it is unwise to treat a particular scheme as the best processor-thread configuration. However, using unthreaded MPI processes is likely to be inefficient as compared to threaded processes. LAMMPS, which does not spawn threads for parallelization, was found to achieve a speed-up that was significantly inferior to that obtained by our hybrid algorithm.

The nano-indentation of a chromium film was performed in such a computational scheme and the results were found to be within expectations. However, the algorithm used for the parallelization is not optimal, and its performance can be enhanced further. There is room of further improvements in the serial algorithm as well.

**APPENDIX**

From the embedded atom method, the total energy of an assembly of atoms is [33]

$$E_t = \sum_i F_i(\rho_i) + \frac{1}{2}\sum_{\substack{i,j \\ i \neq j}} \Phi(r_{ij}) + \sum_i M(P_i) \quad \ldots\ldots\ldots\ldots\ldots(A1)$$

$$\rho_i = \sum_{j(\neq i)} f(r_{ij}) \quad P_i = \sum_{j(\neq i)} f^2(r_{ij}) \quad f(r) = f_e\left(\frac{r_e}{r}\right)^\beta \quad \ldots\ldots\ldots\ldots\ldots(A2)$$

where $E_t$ is the total energy, $\rho$ is the electron density at atom $i$ due to all other atoms, $f(r_{ij})$ is the electron density distribution function of an atom, $r_{ij}$ is the separation distance between atom $i$ and atom $j$ and $r_e$ is the equilibrium interatomic distance, $F(\rho_i)$ is the embedding energy to embed atom $i$ in an electron density $\rho_i$, and $\Phi(r_{ij})$ is the two-body potential between atom $i$ and atom $j$. The analytical term $M(P_i)$ has been introduced to fit the negative Cauchy pressure of Chromium. The functions $F(\rho)$ and $M(P)$ are

$$F(\rho) = -F_o\left[1 - \ln\left(\frac{\rho}{\rho_e}\right)^n\right]\left(\frac{\rho}{\rho_e}\right)^n \quad M(P) = \alpha\left[\left(\frac{P}{P_e} - 1\right)^2\right]\exp\left[-\left(\frac{P}{P_e} - 1\right)^2\right] \quad \ldots\ldots\ldots\ldots\ldots(A3)$$

The pair-potential $\Phi(r)$ proposed by Pasianot [62] has the following form

$$\Phi(x) = (x-d)^2(a_3 x^3 + a_2 x^2 + a_1 x + a_0), \quad x = (r/r_e) \quad x \leq d \ldots\ldots\ldots\ldots\ldots(A4)$$
$$= 0, \quad x > d = 1.65$$

The above model has been applied to the indented film (Cr atoms). The indenter is tetrahedral in shape is composed of fixed carbon atoms in order to simulate the hardness of diamond. The carbon-chromium (C-Cr) interactions, i.e. force on the indenter, have been modeled using a Morse pair potential, where y is the interatomic C-Cr distance.

$$\varphi(y) = D_o\left[\exp\left\{-\alpha_x\left(\frac{y}{y_{eq}} - 1\right)\right\} - 2\exp\left\{-\frac{\alpha_x}{2}\left(\frac{y}{y_{eq}} - 1\right)\right\}\right] \quad \ldots\ldots\ldots\ldots\ldots(A5)$$

The Morse potential parameters were acquired by the lattice inversion method [34]. Incorporating the above equations, we get the following forces:

$$T_{1ij} = -\frac{n^2 F_o \beta f_e}{r \rho_i} \left(\frac{r_e}{r}\right)^\beta \left(\frac{\rho_i}{\rho_{ei}}\right)^n \ln\left(\frac{\rho_i}{\rho_{ei}}\right); \quad r = r_{ij} \quad \text{...............(A6)}$$

$$T_{2ij} = -\frac{4\alpha\beta f_e^2 P_i}{P_{ei}^2 r} \left(\frac{r_e}{r}\right)^{2\beta} \left(2 - \frac{P_i}{P_{ei}}\right)\left(\frac{P_i}{P_{ei}} - 1\right) \exp\left[-\left(\frac{P_i}{P_{ei}} - 1\right)^2\right] \quad \text{...............(A7)}$$

$$T_{3ij} = -\left(\frac{x-d}{r_e}\right)(5a_3 x^3 + (4a_2 - 3a_3 d)x^2 + (3a_1 - 2a_2 d)x + (2a_0 - a_1 d)) \quad x = (r/r_e) \leq d \quad \text{..(A8)}$$

$$T_{4ij} = \frac{D_o \alpha_x}{y_{eq}} \left[\exp\left\{-\alpha_x\left(\frac{y}{y_{eq}} - 1\right)\right\} - \exp\left\{-\frac{\alpha_x}{2}\left(\frac{y}{y_{eq}} - 1\right)\right\}\right] \quad \text{...............(A9)}$$

$$T_i = \sum_{j \neq i, (i,j \in Cr)} (T_{1ij} + T_{2ij} + T_{3ij}) + \sum_{j \neq i, (i \in Cr, j \in C)} T_{4ij} \quad \text{...............(A10)}$$

$T_i$ is the total force on Cr atom(i) by all other chromium atoms. Dividing this by the mass gives the acceleration of atom(i). The various steps of the velocity-Verlet algorithm are shown below.

$$r(t + \Delta t) = r(t) + v(t)\Delta t + \frac{1}{2}a(t)\Delta t^2 \quad \text{...............(A11)}$$

$$v\left(t + \frac{1}{2}\Delta t\right) = v(t) + \frac{1}{2}a(t)\Delta t \quad \text{...............(A12)}$$

$$a(t + \Delta t) = T/m \quad \text{...............(A13)}$$

$$v(t + \Delta t) = v\left(t + \frac{1}{2}\Delta t\right) + \frac{1}{2}a(t + \Delta t)\Delta t \quad \text{...............(A14)}$$